\documentclass[letterpaper]{article} 
\usepackage{amsmath}
\usepackage{amsfonts}
\usepackage{xcolor}
\usepackage{multirow}
\usepackage{aaai23}  
\usepackage{times}  
\usepackage{helvet}  
\usepackage{courier}  
\usepackage[hyphens]{url}  
\usepackage{graphicx} 
\usepackage{natbib}  
\usepackage{caption} 
\usepackage{algorithm2e}
\usepackage{dsfont}
\usepackage{subfigure}

\title{Fast Offline Policy Optimization for Large Scale Recommendation}
\author{
    Otmane Sakhi\textsuperscript{\rm 1, 2},
    David Rohde\textsuperscript{\rm 1},
    Alexandre Gilotte\textsuperscript{\rm 1}
}
\affiliations{
    \textsuperscript{\rm 1}Criteo AI Lab, Paris, France\\
    \textsuperscript{\rm 2}CREST-ENSAE, IPP, Palaiseau, France\\

    \{o.sakhi, d.rohde, a.gilotte\}@criteo.com
}

\begin{document}

\maketitle

\begin{abstract}
Personalised interactive systems such as recommender systems require selecting relevant items from massive catalogs dependent on context. Reward-driven offline optimisation of these systems can be achieved by a relaxation of the discrete problem resulting in policy learning or REINFORCE style learning algorithms. Unfortunately, this relaxation step requires computing a sum over the entire catalogue making the complexity of the evaluation of the gradient (and hence each stochastic gradient descent iterations) linear in the catalogue size. This calculation is untenable in many real world examples such as large catalogue recommender systems, severely limiting the usefulness of this method in practice. In this paper, we derive an approximation of these policy learning algorithms that scale logarithmically with the catalogue size. Our contribution is based upon combining three novel ideas: a new Monte Carlo estimate of the gradient of a policy, the self normalised importance sampling estimator and the use of fast maximum inner product search at training time. Extensive experiments show that our algorithm is an order of magnitude faster than naive approaches yet produces equally good policies. 
\end{abstract}

\section{Introduction}
\label{sec:introduction}
Large Scale Recommender systems are helping users navigate the enormous amount of content present on the internet, allowing them to identify relevant items. From movie recommendation, basket completion to ad placement, all of these systems need to make decisions in an accurate and fast manner. In this work, we cast the problem of recommendation in the offline contextual bandit framework \cite{poem, dr}. Given a context $x$, the decision system performs an action $a$, the context then interacts with the action recommended and we receive a reward $r(a,x)$. We represent the recommender system as a stochastic parametric policy $\pi_\theta: \mathcal{X} \to \mathcal{P(A)}$, which given a context $x \in \mathcal{X}$, defines a probability distribution over the discrete action space $\mathcal{A}$ of size $P$. We suppose that the contexts $x$ are stochastic and coming from an unknown distribution $\nu$ on $\mathcal{X}$. Our objective is to maximize w.r.t to our parameter $\theta$ the average reward over contexts and actions performed by $\pi_\theta$. It can be written as: 
\begin{align}
\label{oracle}
    R(\pi_\theta) &= \mathbf{\mathbb{E}}_{x \sim \nu(\mathcal{X}), a \sim \pi_\theta(.|x)}[r(a, x)]\\
    &= \mathbf{\mathbb{E}}_{x \sim \nu(\mathcal{X})}[\sum_{a \in \mathcal{A}} \pi_\theta(a|x) r(a, x)]
\end{align}

In real world applications, we usually have access to a finite number of context observations $\{x_i\}_{i = 1}^N$ and a reward estimator $\hat{r}(a,x)$ built depending on the application and the task our system is trying to solve. We define an empirical estimator aligned with Equation \eqref{oracle} by:
\begin{align}
\label{objective}
    \hat{R}(\pi_\theta) &= \frac{1}{N}\sum_{i=1}^{N} \mathbf{\mathbb{E}}_{a \sim \pi_\theta(.|x_i)}[\hat{r}(a, x_i)]\\
    &= \frac{1}{N}\sum_{i=1}^{N} \sum_{a \in \mathcal{A}} \pi_\theta(a|x_i) \hat{r}(a, x_i).
\end{align}

This simple equation actually encompasses the majority of the objectives used in the offline bandit literature depending on the reward estimator chosen. For example, if we have access to a bandit dataset with the actions done, their propensities and the reward obtained $\{a_i, p_i, r_i\}_{i = 1}^N$, the IPS/Horwitz-Thompson estimator and its clipped variant \cite{poem, cips} can be obtained by choosing a reward estimator of the following form :
$$ \hat{r}_{IPS}^{\tau}(a, x_i) = 
\begin{cases}
 \frac{r_i}{\max(\tau, p_i)}              & \text{if ~} a = a_i \\
    0 & \text{otherwise }
\end{cases}
$$

\noindent
with $\tau$ the clipping factor between 0 and 1, with the original IPS retreived when $\tau = 0$. With the additional assumption that we have access to a reward model $r_\mathcal{M}$, we can similarly define the Doubly Robust estimator \cite{dr} and its clipped variant \cite{drc} by choosing a reward estimator of the form :
$$ \hat{r}_{DR}^{\tau}(a, x_i) = 
\begin{cases}
 \frac{r_i - r_\mathcal{M}(a, x_i)}{\max(\tau, p_i)} + r_\mathcal{M}(a, x_i)           & \text{if ~} a = a_i \\
    r_\mathcal{M}(a, x_i) & \text{otherwise }.
\end{cases}
$$

One can also cast other methods \cite{switch} into this framework or just optimize any offline metric by designing an adequate reward estimator $\hat{r}$ and plugging it in the simple objective defined in Equation \eqref{objective}. Our method is versatile and can deal with different applications as long as we provide the right reward estimator. In the rest of the paper, we will not make any assumptions on the reward estimator used unless stated explicitly.

\section{Parametrizing the Policy}
\label{inference}

In the problem of recommendation, our objective is to find the best product that matches the current interest of the user. As we deal with distinct enumerable products, we consider our action space discrete and we opt naturally for a policy, that is, conditioned on the context $x$, of the softmax form \cite{poem}:
\[
\pi_\theta(a|x) = \frac{ \exp\{  f_\theta(a, x) \} }{ \sum_b \exp\{ f_\theta(b, x)\}} =  \frac{ \exp\{  f_\theta(a, x) \} }{Z_\theta(x)}
\]

\noindent
with $f_\theta$ a parametric transformation of the context and the action that encodes the relevance of the action $a$ for the context $x$. After training the policy and given a context $x$, the best recommendation online is retrieved by computing the action that maximizes the scores:
\begin{align}
\label{argmax}
a_x^* = {\rm argmax}_a ~ f_\theta(a, x).
\end{align}

This recommendation needs to be done in milliseconds over large catalog sizes making the form of $f_\theta$ crucial to performing Equation \eqref{argmax} rapidly. By restricting the policy to the following form:
\[
f_\theta(a, x) = h_\Xi(x)^T\beta_a
\]

\noindent
with $\theta = [\Xi, \beta]$, $h_\Xi$ a transform that creates a user embedding and $\beta_a$ the item embeddings. Equation \eqref{argmax} becomes:
\begin{align*}
a_x^* = {\rm argmax}_a ~ h_\Xi(x)^T\beta_a
\end{align*}

\noindent
which is precisely the problem that approximate MIPS: Maximum Inner Product Search algorithms \cite{hnsw, faiss} solves quickly (if approximately). This is achieved by first building a fixed index with a particular structure \cite{hnsw} allowing fast identification of the item with the largest inner product with the query $h_\Xi(x)$. With this index, solving Equation \eqref{argmax} is done in a time complexity logarithmic in the the size of the action space $P$ making it possible to deliver recommendations quickly from a large catalog.

\section{Optimizing the Objective}

In large scale recommendation problems, we usually deal with a considerable amount of observations making stochastic gradient descent and its variants \cite{sgd} suitable for such application. As our objective is decomposable, we are interested in the gradient of $\hat{R}_i(\pi_\theta) = \mathbf{\mathbb{E}}_{a \sim \pi_\theta(.|x_i)}[\hat{r}(a, x_i)]$ for a single observation $x_i$. A gradient can be derived using the log trick \cite{REINFORCE}:
\begin{align}
\label{reinforce}
\nabla_\theta \hat{R}_i(\pi_\theta) = \mathbf{\mathbb{E}}_{a \sim \pi_\theta(.|x_i)}[\hat{r}(a, x_i)\nabla_\theta \log \pi_\theta(a|x_i)].
\end{align}

When the action space is of small size \cite{poem, dr, drc}, this gradient can be computed exactly as an expectation over the discrete distribution $\pi_\theta$. Once the size of the catalog $P$ is in the order of millions, an exact gradient update becomes a bottleneck for the optimization process because of the complexity of the following computations:

\noindent
\textbf{1 - Computing $\nabla_\theta \log \pi_\theta(.|x_i)$}: We need to deal with the normalizing constant $Z_\theta(x_i)$ present in the computation of $\nabla_\theta \log \pi_\theta(.|x_i)$. Indeed, $Z_\theta(x_i)$ is a sum over all the action space and its computation needs to be avoided if we hope to reduce the complexity of the gradient update.

\noindent
\textbf{2 - Computing the expectation}: The expectation is a sum over all the action space and is obviously computed in $\mathcal{O}(P)$. To avoid this expensive sum, we can resort to sampling from $\pi_\theta$ to approximate the gradient. This allows us to obtain the REINFORCE estimator \cite{REINFORCE}, an unbiased estimator of the expectation but does not change the complexity of the method which stays linear in the catalog size. Indeed, sampling needs the computation of $Z_\theta(x_i)$ or can be done with the gumbel trick \cite{gumbel} which both scale in $\mathcal{O}(P)$. To lower the time complexity, we need to avoid sampling directly from $\pi_\theta$ and use Monte Carlo techniques instead such as importance sampling \cite{mcbook} with carefully chosen proposals to achieve fast sampling and accurate gradient approximation.

The proposed approach will try to reduce the complexity of the gradient computation by separately dealing  with the issues mentioned above in a principled manner. This will achieve a faster offline training, and will hopefully not suffer a loss in the quality of the policy learned. 

\paragraph{Remark.} The problem we are interested in should not be confused with the maximum log-likelihood problem which is to maximize: 
\begin{align}
\mathcal{L} =  \sum_n  h_\Xi(x_n)^T \beta_{a_n} - \log (\sum_i \exp(h_\Xi(x_n)^T \beta_{i})).
\label{maxlikelihood}
\end{align}
In the context of policy learning, we are seeking a decision rule that maps $x$ to the highest reward action according to $r(a,x)$. This is different from maximising Equation \eqref{maxlikelihood} which enables us to find the model $P(a|x,\Xi)$ that fits the data the most and totally ignores $r(a,x)$.  While both approaches slow down when the catalog size $P$ is very large due to the sum, they are not the same problem. Many existing methods in the literature have been proposed to optimize Equation \eqref{maxlikelihood} when dealing with large action spaces. These include \cite{sampled_softmax, blanc2018adaptive,rawat2019sampled, mikolov2013efficient}. There are some overlaps between the two problems above as both require calculating expectations under a large categorical distribution, but the difference in the loss functions makes the solutions of these problems different as well. For instance, the solution of the policy learning problem is a deterministic policy, putting, conditionally on $x$, all the mass on the action that maximizes $r(a,x)$. If any of the methods suggested to solve the maximum likelihood problem can be adapted to the policy learning case is beyond the scope of this paper.

\section{The Proposed Method}

As pointed out in the previous section, we need a workaround to deal with the presence of the normalizing constant in the gradient. For a fixed observations $x_i$ and similar to the derivations found in \cite{tvo}, we can push further the computation of $\nabla_\theta \log \pi_\theta(a|x_i)$ to obtain a quantity that does not involve $Z_\theta(x_i)$. Indeed, we have for a fixed action $a$:
\begin{align*}
\nabla_\theta \log \pi_\theta(a|x_i) &= \nabla_\theta f_\theta(a, x_i) - \nabla_\theta \log Z_\theta(x_i) \\
&= \nabla_\theta f_\theta(a, x_i) - \frac{\nabla_\theta Z_\theta(x_i)}{Z_\theta(x_i)} \\
&= \nabla_\theta f_\theta(a, x_i) - \sum_b \pi_\theta(b|x_i) \nabla_\theta f_\theta(b, x_i) \\
&= \nabla_\theta f_\theta(a, x_i) -  \mathbf{\mathbb{E}}_{b \sim \pi_\theta(.|x_i)}[\nabla_\theta f_\theta(b, x_i)]
\end{align*}

\noindent
Injecting the above expression of $\nabla_\theta \log \pi_\theta(a|x_i)$ in Equation \eqref{reinforce} leads us to the following covariance gradient \cite{tvo}:
\begin{align}
\nabla_\theta \hat{R}_i(\pi_\theta) = \mathbf{Cov}_{a \sim \pi_\theta(.|x_i)}[\hat{r}(a, x_i),  \nabla_\theta f_\theta(a, x_i)] \label{cov_grad}
\end{align}

\noindent
with $\mathbf{Cov}[A, \boldsymbol{B}] = \mathbf{\mathbb{E}}[(A - \mathbf{\mathbb{E}}[A] ).(\boldsymbol{B} - \mathbf{\mathbb{E}}[\boldsymbol{B}])]$, which is a covariance between $A$ a scalar function, and $\boldsymbol{B}$ a vector. To estimate this, we first estimate the two inner expectations which are then used in estimating the outer
expectation. Note that the covariance in Equation \eqref{cov_grad} is between two deterministic functions of one single random variable $a$ that follows the distribution of $\pi_\theta$. The gradient expression in Equation \eqref{cov_grad} has an intuitive interpretation, as the covariance quantifies how two quantities evolve together, gradient descent using \eqref{cov_grad} will move in directions where the reward and the gradient of the relevance function have the same evolution w.r.t to the action drawn from the policy $\pi_\theta$ enabling our algorithm to favor reward maximization.
 
 The new gradient formula helps us get rid of the normalizing constant present in  $\nabla_\theta \log \pi_\theta(a|x_i)$ but transforms the expectation we had in Equation \eqref{reinforce} into a double expectation; a covariance between the reward estimator and the gradient of our relevance function $f_\theta$. By exploiting this identity, we remove the requirement to compute $Z_\theta(x_i)$ which scales linearly with the catalogue size if we have available to us two (or more) samples from the policy $\pi_\theta(.|x_i)$ (computing covariances requires multiple samples). Unfortunately, sampling from $\pi_\theta(.|x_i)$ also scales in $\mathcal{O}(P)$ so it seems that no progress has been made.
 
 Wanting to avoid sampling from the current policy $\pi_\theta$, we use self normalized importance sampling \cite{mcbook} to approximate the expectations without relying on the computation of the normalizing constant $Z_\theta$. Indeed, for a fixed $x_i$, if we have access to a discrete proposal $q$ over the action space, one can build an estimator of the expectation of a general function $g$ under $\pi_\theta(.|x_i)$ by: $$\mathbf{\mathbb{E}}_{\pi_\theta(.|x_i)}[g(a)] \approx \sum_{s = 1}^S \Bar{\omega}_s g(a_s)$$
 
 \noindent
 with $a_s \sim q \quad \forall s$, $\omega_s = \frac{\exp \{f(a_s, x_i)\}}{q(a_s)}$ and $\Bar{\omega}_s = \frac{\omega_s}{\sum_{s' = 1}^S \omega_{s'}}$.
 
This algorithm removes the dependency on the catalogue
size by avoiding the computation of $Z_\theta(x_i)$.
The cost for this is that the estimator is now biased with a bias decreasing with how close the proposal $q$ is to the policy $\pi_\theta$ \cite{chi2}. This means that if we want to exploit self normalized importance sampling efficiently, we need to have access to a proposal $q$ that is \textbf{fast} to sample from, easy to \textbf{evaluate} as its needed to compute the weights $\Bar{\omega}_s$ and \textbf{close} to the actual policy $\pi_\theta$ to reduce the bias and the variance of the method. To build such proposal $q$ that respects the three conditions, we need an additional assumption on the parameters learned of our policy $\pi_\theta$.

\paragraph{Assumption 1:} The item embedding matrix $\beta$ is fixed and we are only interested in learning the user embedding transform $h_\Xi$, which means  $\Xi =\theta$.\\

Making this assumption in the context of recommendation is reasonable \cite{meanemb}. We can learn different representations of the items from collaborative filtering data, text data or even images making learning meaningful embeddings possible without relying on the downstream task we are trying to solve.

This assumption allows us to fully exploit approximate MIPS algorithms in the training phase by building an index over the item embeddings $\beta$ and fixing it before beginning the optimization of our policy. Indeed, while we can modify existing embeddings in the index for a logarithmic complexity \cite{hnsw} in the training phase, this procedure will further slow down the method as we need to update the items in the index for each training iteration. 

Making \textbf{Assumption 1} simplifies drastically the procedure as $\beta$ are considered fixed and we only need to compute the index once before the start of the training of our policy. With the help of the approximate MIPS index, and for any $x_i$, we define the proposal $q$ as a mixture between a distribution over the $K$ most probable actions under $\pi_\theta(.|x_i)$ retrieved by the approximate MIPS algorithm i.e $\alpha_K(x_i)= {\rm argsort}(h_{\theta}(x_i)^T\beta)_{1:K}$ and a uniform distribution over actions. It can be expressed as:
\[
    q_{K, \epsilon}(a|x_i)= 
\begin{cases}
    \frac{\epsilon}{P} + (1-\epsilon)\kappa(a|x_i),              & \text{if ~} a \in \alpha_K(x_i) \\
    \frac{\epsilon}{P},& \text{otherwise }.
\end{cases}
\]

Where $\epsilon$ is a parameter that controls the mixture and 
\[
\kappa(a|x_i) = \frac{\exp(h_{\theta}(x_i)^T\beta_a)}{\sum_{a' \in \alpha_K(x_i)}  \exp(h_{\theta}(x_i)^T\beta_{a'})}\mathds{1}[a \in \alpha_K(x_i)].
\]

This proposal answers the necessary conditions to make self normalised importance sampling works efficiently:
\begin{itemize}
    \item It is \textbf{fast} to sample from as a mixture of a uniform distribution and a distribution constructed with approximate MIPS making the time complexity $\mathcal{O}(\log P)$ \cite{hnsw}. Indeed, solving the argsort, thus constructing $\alpha_K$ can be done logarithmically in the catalog size with the help of approximate MIPS. 
    \item Easy to \textbf{evaluate} as once we have the set $\alpha_K$, the computation will require at maximum a sum over the top K retrieved actions with $K \ll P$.
    \item \textbf{Close} to $\pi_\theta$ as it exploits information about the top actions under $\pi_\theta$ and covers well the early stage (when $\pi_\theta$ close to uniform) and late stage (when $\pi_\theta$ is degenerate on the top actions) of the optimization process.
\end{itemize}

\RestyleAlgo{ruled}
\begin{algorithm}
\caption{Fast Offline Policy Learning}\label{alg:one}
\textbf{Inputs:} D = $\{x_i\}_{i = 1}^N$, \text{reward estimator } $\hat{r}$, the item embeddings $\beta$\\

\textbf{Parameters:} $T \ge 1, \alpha, K, S \ge 2, \epsilon \in [0, 1]$ \\
\textbf{Initialise:} $\theta = \theta_0$, approximate MIPS index for $q_{\epsilon, K}$ \\
\For{$t = 0$ \textbf{to} $T$} 
{
$x \sim D$ \\
\text{query $h_\theta(x)$ to get the approx. top $K$ actions set $\alpha_K$}\\
\text{build the proposal} $q_{\epsilon, K}(.|x)$ \\
sample $a_1, ..., a_S \sim q_{\epsilon, K}(.|x)$ \\
\textbf{Estimate the covariance gradient:}\\
$\hat{r}_s = \hat{r}(a_s,x), \nabla f_s = \nabla_\theta f_\theta(a_s, x) \quad \forall s $\\
$\omega_s \gets \frac{\exp\{ f(a_s, x)\}}{q_{\epsilon, K}(a_s|x)}, \Bar{\omega}_s \gets \frac{\omega_s }{\sum_s \omega_s} \quad \forall s$ \\
$\Bar{r} \gets \sum_{s = 1}^S \Bar{\omega}_s \hat{r}_s$,  $\Bar{\nabla f} \gets \sum_{s = 1}^S \Bar{\omega}_s \nabla f_s$ \\
$grad_\theta \gets \sum_{s = 1}^S \Bar{\omega}_s [\hat{r}_s - \Bar{r}][\nabla f_s - \Bar{\nabla f}]$ \\
\textbf{Update the policy parameter} $\theta$: \\
$\theta \gets \theta - \alpha grad_\theta$
}
\textbf{return} $\theta$ 
\end{algorithm}
The performance of the self normalized importance sampling algorithm using our proposal is controlled by the number of Monte Carlo samples $S$, the mixture parameter $\epsilon$ and
the number of items returned by the maximum inner product
search $K$. This performance can also be impacted by the parameters of the approximate MIPS algorithms that trade-off speed of retrieval for the accuracy of the argsort, changing the parameters seemed to have little to no impact on the study so we decided to fix them for the whole experiments.

By combining the self normalized importance sampling algorithm with our mixture proposal, a stochastic gradient descent \cite{sgd} version of the algorithm we suggest can be described in Algorithm \ref{alg:one}.

This procedure is compatible with any stochastic first order optimization algorithm \cite{sgd, adam}. In the next section, we will validate the approach on real world datasets and study the impact of our method on the training speed and the quality of the learned policy.

\section{Experimental Results}

We test our approach on a session completion task using a collaborative filtering dataset. In the training phase, we split randomly the user-item interaction session into two parts, $X$ and $Y$. $X$ is used as observed and we will condition on it to predict items that are found in $Y$, or in other words, predict items that complements the vector $X$. Our goal is to build an algorithm that given a user-item interaction vector $X_{new}$, predict or recommend items that may interest the user. This can be cast into an offline bandit framework where our policy $\pi_\theta$ takes the observed part $X$ as a context, recommends an item $a$ and receives the binary reward $\hat{r}(a,X) = \mathds{1}[a \in Y]$. The goal then is to learn a policy $\pi_\theta$ that will maximize the reward $\hat{r}$ allowing it to solve the session completion task. Note that our method is very versatile and can deal with different problems as long as they can be cast into an offline policy learning problem as described in the introduction. We chose a session completion task for simplicity as user-item interaction datasets are public making the experiments easily reproducible.

To prepare our experiment, we begin by splitting the user-item sessions into two independent sessions with the same number of interactions: the observed items $X$ and the complementary items $Y$. Once this split performed, we also split the whole dataset into a train $D_{train} = [X_{train}, Y_{train}]$ and test split $D_{test} = [X_{test}, Y_{test}]$.  Given the train split $D_{train}$, we use the observed contexts $X_{train}$ to first compute the item embeddings $\beta$ using SVD matrix decomposition \cite{svd} of dimension $[P,L]$ with $L \ll P$ that will be fixed to create the approximate MIPS index. This index is then used by our proposal $q$ in the learning phase, but also used to get the best recommendation rapidly in the testing phase as described in Equation \eqref{argmax}.

\begin{table}
\centering
\begin{tabular}{ |c||c|c|  }
 \hline
 & Catalog size & Number of users \\
 \hline
 \textbf{Twitch}  & 790K & 500K \\
 \textbf{GoodReads}&   2.33M  & 300K \\
 \hline
\end{tabular}
\caption{The statistics of the datasets after processing}
\label{tab:stats}
\end{table}

Once we have the item embedding matrix $\beta$, we compute the user contexts $x$ as the mean embeddings \cite{meanemb} of the items that the user interacted with. For $X_i$ the observed item interactions of user $i$ and $n_i$ the number of items the same user interacted with, we define  $x_i^{emb} = \frac{1}{n_i} \sum_{a \in X_i} \beta_a$. The obtained vector $x_i^{emb}$ is of dimension $L$ and will be the user context in our experiment, meaning that we will use $D_{train}^{emb} = [x^{emb}_{train}, Y_{train}]$ and $D_{test}^{emb} = [x^{emb}_{test}, Y_{test}]$ instead of $D_{train}$ and $D_{test}$. 

\begin{figure*}
  \centering
 \includegraphics[scale=0.345]{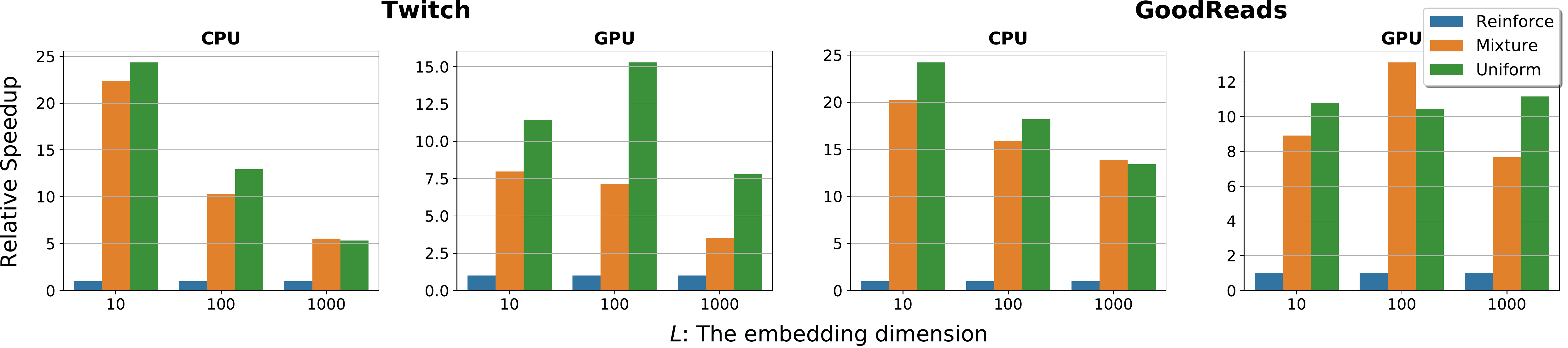}
    \caption{The speedups of the proposed algorithms on the Twitch and GoodReads datasets, with and without GPU acceleration.}
\label{all_speed}
\end{figure*}

The next step is to parametrize the policy $\pi_\theta$ that we will train using our algorithm. With the item embeddings $\beta$ fixed, we only need to define the user transform $h_\theta$. We take $h_\theta$ to be a linear function of the user context $x^{emb}$ defined above ie. $h_\theta(x^{emb}) = \theta^Tx^{emb}$ making the dimension of the parameter learned $\theta$ equals to $[L, L]$. After training the policy $\pi_\theta$, we validate its performance on the test split $D^{emb}_{test}$ by computing the average reward collected after querying the argmax of our policy using approximate MIPS ie.
\begin{align*}
    R_{test} &= \frac{1}{N_{test}} \sum_{j = 1}^{N_{test}} \mathds{1}[{\rm argmax}_a f_\theta(a, x^{emb}_j) \in Y_j]
\end{align*}
We choose two collaborative filtering datasets with large catalog sizes to validate our approach. the Twitch dataset \cite{twitch} and the GoodReads user-books interaction dataset \cite{gr1, gr2} with both representing a good test bed for the scalability of the proposed methods. We transform the datasets into a user-item interaction matrix with statistics represented in Table \ref{tab:stats}. To build the approximate MIPS index, we use the HNSW algorithm \cite{hnsw} bundled in the FAISS library \cite{faiss}. The optimization routine is implemented using Pytorch \cite{torch}, and  we opt for the Adam optimizer \cite{adam} with a batch size of 32 and a learning rate of $10^{-4}$ for all the experiments with the twitch dataset and $5.10^{-5}$ for the goodreads dataset. The source code\footnote{https://github.com/criteo-research/fopo} to reproduce the results has all implementation details.

With the experimental protocol described above, we want to study the speed improvement brought by our method, and the effect of the parameters controlling our proposal $q_{\epsilon, K}$, namely the mixture parameter $\epsilon$ and the number of top $K$ retrieved items on the speed and the quality of the policy learned. The rest of the experiments section will be decomposed into different research questions that we would like to answer for a better understanding of the approach proposed. 

\noindent
\textbf{RQ0 - What is the cost of fixing the embeddings?} 

To verify the impact of making \textbf{Assumption 1}, we compare the performance and speed of REINFORCE with the item embeddings $\beta$ fixed to REINFORCE with $\beta$ initialised with the SVD decomposition and trained. We report on Table \ref{tab:notfixing}, for both datasets and two different values of the embedding dimension $L \in \{10, 100\}$, the relative speedup $rS = T_{\textit{trained}}/T_{\textit{fixed}}$ and the relative performance $rP = R_{\textit{trained}}/R_{\textit{fixed}}$ with $T_{\textit{method}}$ and $R_{\textit{method}}$ are respectively the run time and reward of a method after training.

We observe that training $\beta$ have no to little impact on the training time, but hurts the performance of REINFORCE mainly due to the variance introduced by having more parameters to optimize. Indeed, we observe that increasing the dimension $L$ worsens the relative performance $rP$ in Table \ref{tab:notfixing}. We conclude that in our experiments, \textbf{Assumption 1} can be made as it does not negatively affect the training.

\begin{table}
\centering
\begin{tabular}{ |c||c|c|c|  }
 \hline
 Dataset & $L$  & $rP$ & $rS$ \\
 \hline
 \multirow{2}{*}{\textbf{Twitch}} & 10 & \textit{0.54} & \textbf{1.01} \\
    & 100 & \textit{0.40} & \textbf{1.01}\\
    \hline
 \multirow{2}{*}{\textbf{GoodReads}} & 10 & \textit{0.83} & \textbf{1.00}\\
    & 100 & \textit{0.71} & \textbf{1.01}\\
    \hline
\end{tabular}
\caption{Impact of fixing the product embeddings.}
\label{tab:notfixing}
\end{table}
\noindent
\textbf{RQ1 - What is the speed gain over REINFORCE?} 

This work addresses the computational inefficiency of vanilla REINFORCE as we previously described. To show that our proposed method is a good solution to the problem presented, we conduct extensive experiments to study the acceleration brought by our methods relatively to our baseline. In this section, we quantify the gains in form of a relative speedup of the proposed methods compared to REINFORCE; for the same experimental setup, if $T_{\text{method}}$ is the wall time of a method, we define the relative speedup to our baseline as $RS_{\text{method}} = T_{\text{REINFORCE}}/T_{\text{method}}$. 

Before we dive into the experiments, we first identify the different parameters that can have a big impact on the running time of the learning algorithm. As Matrix-vector multiplication is naively done in $\mathcal{O}(L^2)$ with $L$ being the embedding dimension, we conduct multiple experiments on the datasets we have at our disposal, on \textbf{CPU} and \textbf{GPU} devices while changing the embedding dimension $L$ to have a good understanding of the speed gains in different settings. In Figure \ref{all_speed}, we compare REINFORCE, the algorithm that approximates the gradient expression in Equation \eqref{reinforce} by samples from the true policy, with the proposed approach for $\epsilon = 1$ for which our proposal becomes uniform, and the mixture algorithm with $\epsilon \neq 1$, represented in these experiments by a run with $\epsilon = 0.8$. Note that the value of $\epsilon$ does not affect the running time as long as it is different from 1. In these experiments, we fix $K = 256$ and $S = 1000$.

\begin{figure*}
  \centering
 \includegraphics[scale=0.345]{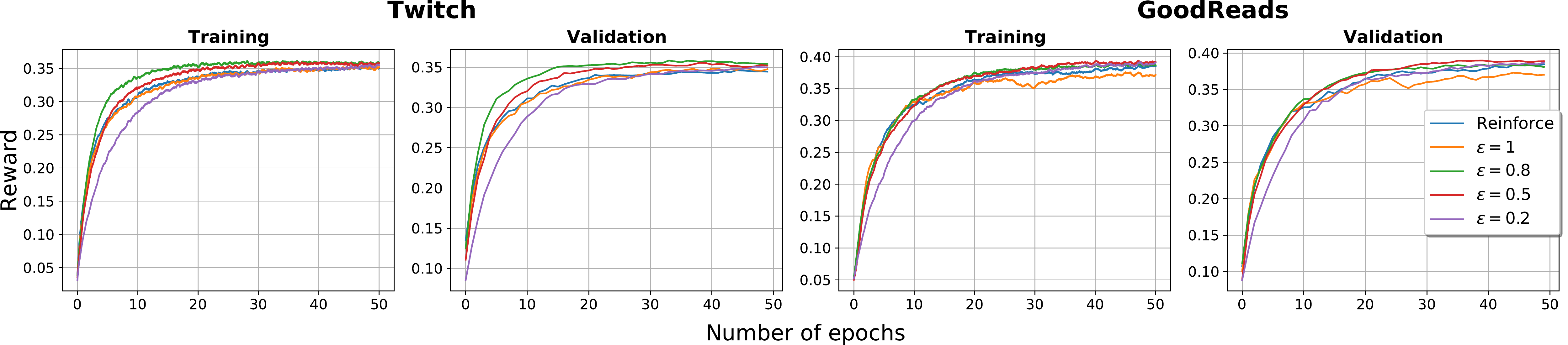}
    \caption{The performance of the algorithms while changing the mixture parameter $\epsilon$ on both Twitch and GoodReads datasets.}
\label{all_perf}
\end{figure*}

We can see from Figure \ref{all_speed} that we gain significant speedups in the different settings considered ranging from 5 to 25 faster offline policy training compared to REINFORCE, with the largest speedups observed on CPU machines (no massive parallelization contrary to GPUs) and with small embedding dimension $L$ as the complexity of matrix-vector multiplication $\mathcal{O}(L^2)$ does not dominate the gradient update complexity. We can see that even if we have access to a GPU, and we make $L$ big enough to learn better user embeddings, the speedup is still interesting as we still obtain 5-10 times speedup compared to our baseline. Note that all the run times were averaged over 5 epochs, and that these time comparisons can differ depending on the computational resources at ones disposal, but we argue that the differences should be more important when training on \textbf{CPU} machines, especially when the user embeddings are easy to compute, making offline policy training on less powerful machines possible. We also expect bigger gains when dealing with larger action spaces given that our method scales better with growing catalogs.  

\noindent
\textbf{RQ2 - What is the impact of changing $\epsilon$?} 

We have seen that changing $\epsilon$, especially setting it to $1$ can have a significant impact on the iteration cost of the optimization procedure. To get a better understanding of our method, we need to study the effect of the mixture parameter on the the quality and thus the performance of the trained policies. In this section, we investigate the impact of changing $\epsilon$ on the average reward, on the same problem with an embedding dimension set to $L = 1000$ to make the learning problem difficult, while also fixing the other parameters to $K = 256$ and $S = 1000$. The algorithms were run on the datasets considered for 50 epochs, and \textbf{GPU} training was used to be fair to the baseline as the speedup gains in this particular setting are the lowest. We plot the results of these runs on Figure \ref{all_perf}. We observe that, even if REINFORCE has a much bigger time complexity per iteration (scales linearly on the catalog size), it does not outperform the optimization routines suggested by our approach. Indeed, we can achieve the same level of performance, sometimes performances beyond what REINFORCE can reach with much faster training, so not only our method is faster than REINFORCE, it can also lead to a better optima as we suspect that using the index on the training phase helps the policies be better aligned with how the recommendation is done after deployment, as the same index is used online.   

From the same plots, we can also conclude that fixing $\epsilon = 1$, even if it provides the fastest approximation as it boils down to using a uniform proposal, is far from being optimal if our main goal is to obtain the policy with the best average reward. Indeed, the optimal policy is obtained with values of $\epsilon \neq 1$ ($\epsilon = 0.8$ for the Twitch Dataset and $\epsilon = 0.5$ for the Goodreads dataset). Even if this means that we need to try out different values of $\epsilon$ to get the best out of our approach, it confirms that our first intuition of building a mixture proposal between the uniform distribution and a TOP-$K$ distribution brings value, in terms of iteration speed and quality of the policy learned as our proposal is expected to behave well in all training phases; $\epsilon \approx 1$ is expected to work well in the beginning of the optimization while $\epsilon$ close to 0 is expected to work when the policy is close to convergence.

Figure \ref{all_perf} plot the evolution per epoch of the performance of the policies trained with the different algorithms. As the cost of iteration significantly changes depending on the algorithm used, we might be interested in a comparison given a fixed time budget, allowing us to train the policy with different methods for the same amount of time.

\begin{figure}
    \centering
\includegraphics[scale=0.325]{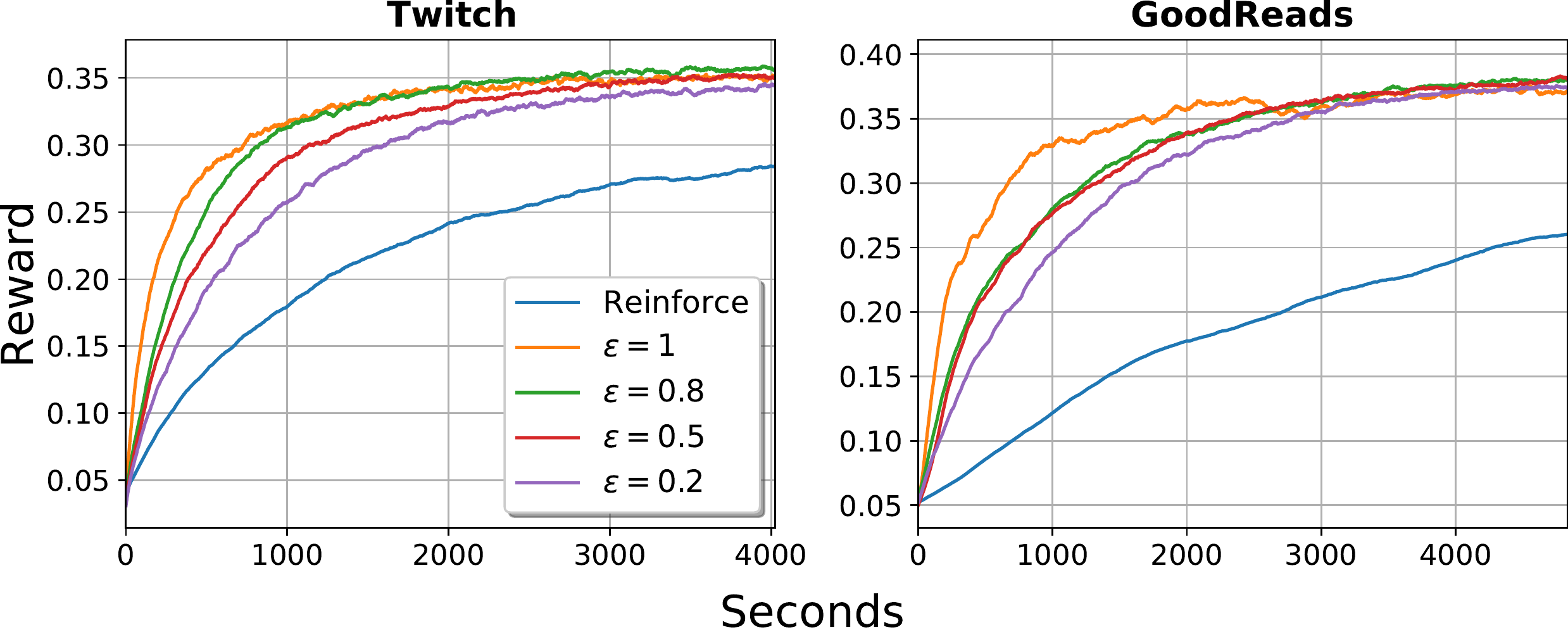}
  \label{fig:val_sec}
    \caption{Training performance given a fixed time budget.}
\label{time_perf}
\end{figure}

We provide Figure \ref{time_perf} to shed more light on how the different algorithms perform on the training phase with a fixed time budget. As the uniform proposal provides the fastest training, we consider its running time after 50 epochs the allowed time budget and compare the performance of all the other methods to it given that fixed running time. We can observe that REINFORCE is the worst behaving algorithm as its iteration cost scales linearly with $P$ and that a well chosen mixture have always a slight edge over the uniform proposal on both training datasets.  

\noindent
\textbf{RQ3 - What is the impact of changing the number of the top retrieved items $K$?}  

\begin{figure}
  \centering
  \includegraphics[width=\linewidth]{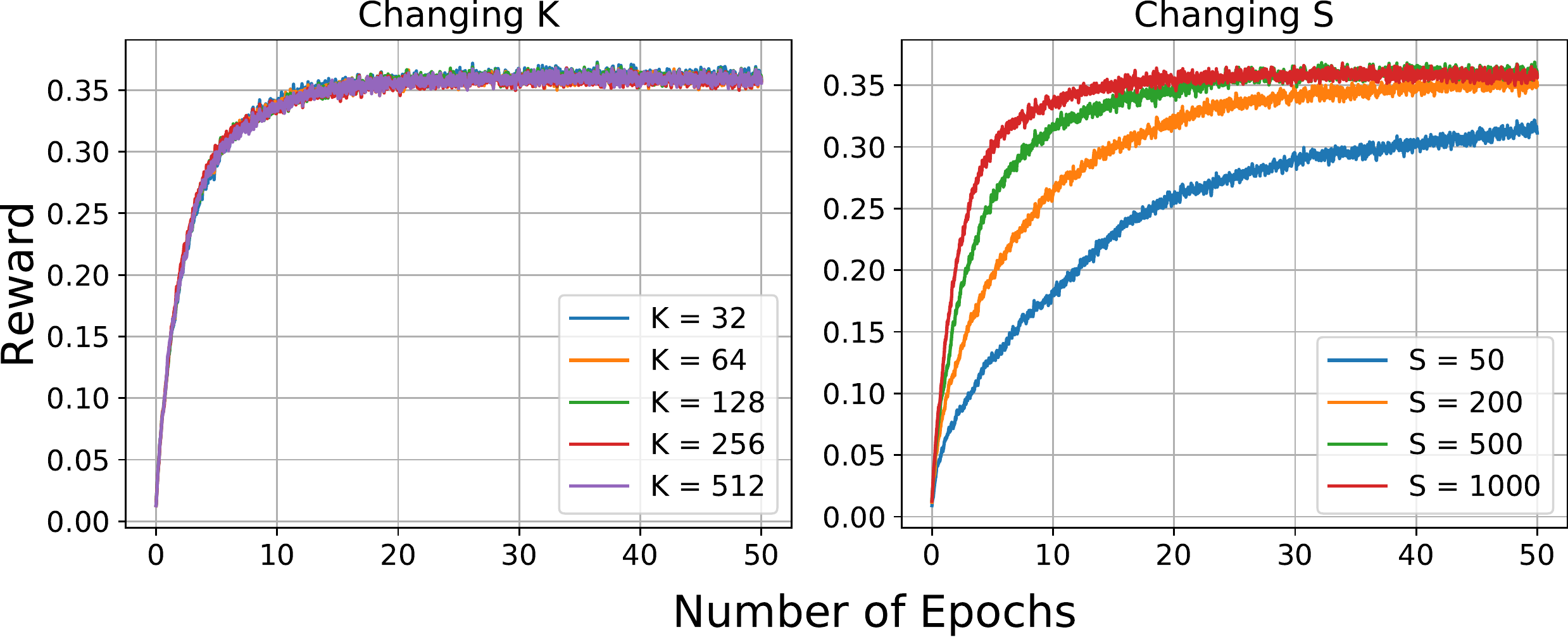}
\captionof{figure}{The effect of changing $K$ and $S$ on the training.}
\label{change}
\end{figure}

As our approach have different hyperparameters, we want to understand how changing them affect the behavior of the algorithms proposed. To quantify the impact of $K$, we focus on the Twitch dataset and fix $\epsilon$ to 0.8, $L = 1000$ and $S = 1000$. We then try different values for  $K \in \{32, 64, 128, 256, 512 \}$, the number of top items returned by approximate MIPS. We run the optimization for 50 epochs and plot the reward of the policies on the training phase in Figure \ref{change}. We observe that the performance is robust to the choice of $K$ as long as its selected big enough to cover most of the top candidates. Even if it is not apparent on the plot, we can also confirm that the iteration cost is not greatly influenced by the choice of $K$ as long as it is orders of magnitude smaller than the catalog size.  

\noindent
\textbf{RQ4 - What is the impact of changing the number of Monte Carlo samples $S$?}  

The number of Monte Carlo samples $S$ controls the approximation accuracy of the gradient as increasing it will reduce the bias (when using the covariance gradient with our proposal $q_{\epsilon, K}$) and the variance of our gradient estimate for an additional computation cost. To understand the impact of $S$ on the training, we restrict ourselves to the Twitch dataset and use $q_{\epsilon = 0.8, K}$ as a proposal. We fix $L = 1000$, $K = 256$ and plot the results of changing $S \in \{50, 200, 500, 1000\}$ in Figure \ref{change}. We run the optimization for 50 epochs. We observe that our policies converge to a better optima when we increase $S$, and that is expected as less noise is present in the gradient approximation \cite{mcbook}. We also observed from our experiments that the average run time only increased slightly from $S = 50$ to $S = 1000$, suggesting that increasing $S$ to a higher value ($S \ll P$) is beneficial for training policies offline as the additional computational cost on \textbf{GPU} is minimal compared to the convergence speed gains achieved.

\section{Related Work}

There has been a surge of interest in off-policy learning in the last decade.  Most of these contributions have focused upon improving the estimate of the reward function including the development of estimators for the reward representing sophisticated modifications of the estimators $\hat{r}_{IPS}$ and $\hat{r}_{DR}$ introduced in the first section.  Correctly estimating the reward is an extremely difficult problem and naive estimators suffer from very high variance
especially when the new policies we want to evaluate are far from the logging policy \cite{chi2}. Many new methods improved the bias-variance trade-off of the reward estimators \cite{cips, dr, drc, switch, snips} and tackled the learning problem by using these estimators within the framework of Empirical Risk Minimization \cite{cips, dr, drc}, or by leveraging refined statistical learning techniques, giving birth to Sample Variance Penalization \cite{poem}, Distributionally Robust Counterfactual Risk Minimization \cite{dro1, dro2}, PAC-Bayesian Counterfactual Risk Minimization \cite{pacbayes} and Imitation Offline Learning \cite{iml}.  Alternatively, instead of using an estimator, an explicit reward model can be used \cite{blob, pess_reward}.  Our contribution is orthogonal to these developments as we consider the problem of how to efficiently find a good policy given a reward function.  Most past work assumes the action space is small and hence the optimization problem is tractable.  Our algorithm reduces the optimization cost associated with large action spaces but the estimation challenges of the reward function remain.

Policy learning emerges as the optimisation problem of a reward driven recommender system. Recommender system training is also sometimes formulated as a maximum likelihood approach when training a model  \cite{vae-cf, esae} to predict missing entries (e.g. predicting missing elements of the MovieLens dataset \cite{movielens}).  Maximum likelihood estimation also suffers from a computational cost that scales in $\mathcal{O}(P)$ but the problem has a different mathematical form to policy learning and methods developed in the maximum likelihood context \cite{negative_sampling, nce, bpr} cannot be applied to policy learning.

There has been limited attention in the literature with regards to scaling offline policy learning methods to the problems of recommendations with large discrete action spaces. For example \cite{rl-large} considered the acceleration of reinforcement learning \cite{rl} which is an online learning framework by definition. Their proposed approaches uses Fast K-NN \cite{hnsw}  algorithms to generate action candidates for the critic to choose from, in contrast to our approach which deals with offline learning and uses approximate MIPS algorithms to define a proposal to better estimate the gradient.

Recently \cite{minminchen} showed that offline policy learning methods can perform well on large scale production systems, introducing a correction to the REINFORCE gradient with little focus on the scalability of the method. Our work addresses the computational issues linked to offline policy learning making it fast to achieve without deteriorating the performance of the obtained policy.

\section{Conclusion and Future Work}

Offline Policy learning is a powerful paradigm for recommender system training as it seeks to align the offline optimization problem with the real world reward measured at A/B test time.  Unfortunately, the $\mathcal{O}(P)$ cost of training traditional policy learning algorithms has limited the widespread application of these methods in large scale decision systems when $P$ is often very large. To deal with this issue, we introduced an efficient offline optimization routine for softmax policies, tailored for the problem of large scale recommendation. Our algorithm makes the training orders of magnitude faster (up to 30x faster in our experiments).  The quality of our policies were at least as good as those found with slower baselines. 
This work can enable practitioners to explore policy learning methods when dealing with large action spaces without relying on huge computational resources. We hope that the solution provided in this paper is a first step towards the adoption of offline policy methods for large scale recommender systems.

There are a number of avenues of open research.  Self normalized importance sampling produces biased gradients that can affect the convergence of stochastic gradient descent, convergence in this setting is not well studied apart from some special cases \cite{rm_sgd, bsgd}.  We showed that fixing $\epsilon$ works well in our experiments, but maybe an adaptive $\epsilon$ may work better still.  We would expect $\epsilon=0$ to work well in the early stages of training and $\epsilon \rightarrow 1$ in the late stages, but robustly determining how to evolve $\epsilon$ is unclear.
Finally, we would also want to explore ways to enable fast training without fixing the item embeddings $\beta$.

\section*{Acknowledgments}
The authors are grateful for all the fruitful discussions engaged with Clément Calauzènes and Nicolas Chopin which
helped improve this work. We would also want to thank the reviewers for their valuable feedback.

\bibliography{bibfile}

\begin{thebibliography}{42}
\providecommand{\natexlab}[1]{#1}

\bibitem[{Agapiou et~al.(2017)Agapiou, Papaspiliopoulos, Sanz-Alonso, and
  Stuart}]{chi2}
Agapiou, S.; Papaspiliopoulos, O.; Sanz-Alonso, D.; and Stuart, A.~M. 2017.
\newblock Importance sampling: Intrinsic dimension and computational cost.
\newblock \emph{Statistical Science}, 405--431.

\bibitem[{Bengio and Senecal(2003)}]{sampled_softmax}
Bengio, Y.; and Senecal, J.-S. 2003.
\newblock Quick Training of Probabilistic Neural Nets by Importance Sampling.
\newblock In Bishop, C.~M.; and Frey, B.~J., eds., \emph{Proceedings of the
  Ninth International Workshop on Artificial Intelligence and Statistics},
  volume~R4 of \emph{Proceedings of Machine Learning Research}, 17--24. PMLR.
\newblock Reissued by PMLR on 01 April 2021.

\bibitem[{Blanc and Rendle(2018)}]{blanc2018adaptive}
Blanc, G.; and Rendle, S. 2018.
\newblock Adaptive sampled softmax with kernel based sampling.
\newblock In \emph{International Conference on Machine Learning}, 590--599.
  PMLR.

\bibitem[{Bottou et~al.(2013)Bottou, Peters, Qui{{\~n}}onero-Candela, Charles,
  Chickering, Portugaly, Ray, Simard, and Snelson}]{cips}
Bottou, L.; Peters, J.; Qui{{\~n}}onero-Candela, J.; Charles, D.~X.;
  Chickering, D.~M.; Portugaly, E.; Ray, D.; Simard, P.; and Snelson, E. 2013.
\newblock Counterfactual Reasoning and Learning Systems: The Example of
  Computational Advertising.
\newblock \emph{Journal of Machine Learning Research}, 14(65): 3207--3260.

\bibitem[{Chen et~al.(2019)Chen, Beutel, Covington, Jain, Belletti, and
  Chi}]{minminchen}
Chen, M.; Beutel, A.; Covington, P.; Jain, S.; Belletti, F.; and Chi, E.~H.
  2019.
\newblock Top-K Off-Policy Correction for a REINFORCE Recommender System.
\newblock In \emph{Proceedings of the Twelfth ACM International Conference on
  Web Search and Data Mining}, WSDM '19, 456–464. New York, NY, USA:
  Association for Computing Machinery.
\newblock ISBN 9781450359405.

\bibitem[{Chopin and Papaspiliopoulos(2020)}]{mcbook}
Chopin, N.; and Papaspiliopoulos, O. 2020.
\newblock \emph{Importance Sampling}, 81--103.
\newblock Cham: Springer International Publishing.
\newblock ISBN 978-3-030-47845-2.

\bibitem[{Dud{\'\i}k et~al.(2014)Dud{\'\i}k, Erhan, Langford, and Li}]{dr}
Dud{\'\i}k, M.; Erhan, D.; Langford, J.; and Li, L. 2014.
\newblock Doubly robust policy evaluation and optimization.
\newblock \emph{Statistical Science}, 29(4): 485--511.

\bibitem[{Dulac-Arnold et~al.(2015)Dulac-Arnold, Evans, van Hasselt, Sunehag,
  Lillicrap, Hunt, Mann, Weber, Degris, and Coppin}]{rl-large}
Dulac-Arnold, G.; Evans, R.; van Hasselt, H.; Sunehag, P.; Lillicrap, T.; Hunt,
  J.; Mann, T.; Weber, T.; Degris, T.; and Coppin, B. 2015.
\newblock Deep reinforcement learning in large discrete action spaces.
\newblock \emph{arXiv preprint arXiv:1512.07679}.

\bibitem[{Faury et~al.(2020)Faury, Tanielian, Vasile, Smirnova, and
  Dohmatob}]{dro1}
Faury, L.; Tanielian, U.; Vasile, F.; Smirnova, E.; and Dohmatob, E. 2020.
\newblock Distributionally Robust Counterfactual Risk Minimization.
\newblock In \emph{AAAI}.

\bibitem[{Gutmann and Hyv\"arinen(2010)}]{nce}
Gutmann, M.; and Hyv\"arinen, A. 2010.
\newblock {N}oise-contrastive estimation: {A} new estimation principle for
  unnormalized statistical models.
\newblock In Teh, Y.; and Titterington, M., eds., \emph{Proc. Int. Conf. on
  Artificial Intelligence and Statistics (AISTATS)}, volume~9 of \emph{JMLR
  W\&CP}, 297--304.

\bibitem[{Harper and Konstan(2015)}]{movielens}
Harper, F.~M.; and Konstan, J.~A. 2015.
\newblock The MovieLens Datasets: History and Context.
\newblock \emph{ACM Trans. Interact. Intell. Syst.}, 5(4).

\bibitem[{Hsieh, Mertikopoulos, and Cevher(2021)}]{bsgd}
Hsieh, Y.-P.; Mertikopoulos, P.; and Cevher, V. 2021.
\newblock The Limits of Min-Max Optimization Algorithms: Convergence to
  Spurious Non-Critical Sets.
\newblock In Meila, M.; and Zhang, T., eds., \emph{Proceedings of the 38th
  International Conference on Machine Learning}, volume 139 of
  \emph{Proceedings of Machine Learning Research}, 4337--4348. PMLR.

\bibitem[{Huijben et~al.(2021)Huijben, Kool, Paulus, and van Sloun}]{gumbel}
Huijben, I.~A.; Kool, W.; Paulus, M.~B.; and van Sloun, R.~J. 2021.
\newblock A Review of the Gumbel-max Trick and its Extensions for Discrete
  Stochasticity in Machine Learning.
\newblock \emph{arXiv preprint arXiv:2110.01515}.

\bibitem[{Jeunen and Goethals(2021)}]{pess_reward}
Jeunen, O.; and Goethals, B. 2021.
\newblock \emph{Pessimistic Reward Models for Off-Policy Learning in
  Recommendation}, 63–74.
\newblock New York, NY, USA: Association for Computing Machinery.
\newblock ISBN 9781450384582.

\bibitem[{Johnson, Douze, and J{\'e}gou(2019)}]{faiss}
Johnson, J.; Douze, M.; and J{\'e}gou, H. 2019.
\newblock Billion-scale similarity search with {GPUs}.
\newblock \emph{IEEE Transactions on Big Data}, 7(3): 535--547.

\bibitem[{Kingma and Ba(2014)}]{adam}
Kingma, D.~P.; and Ba, J. 2014.
\newblock Adam: A method for stochastic optimization.
\newblock \emph{arXiv preprint arXiv:1412.6980}.

\bibitem[{Klema and Laub(1980)}]{svd}
Klema, V.; and Laub, A. 1980.
\newblock The singular value decomposition: Its computation and some
  applications.
\newblock \emph{IEEE Transactions on Automatic Control}, 25(2): 164--176.

\bibitem[{Koch et~al.(2021)Koch, Benhalloum, Genthial, Kuzin, and
  Parfenchik}]{meanemb}
Koch, O.; Benhalloum, A.; Genthial, G.; Kuzin, D.; and Parfenchik, D. 2021.
\newblock Scalable representation learning and retrieval for display
  advertising.
\newblock \emph{arXiv preprint arXiv:2101.00870}.

\bibitem[{Liang et~al.(2018)Liang, Krishnan, Hoffman, and Jebara}]{vae-cf}
Liang, D.; Krishnan, R.~G.; Hoffman, M.~D.; and Jebara, T. 2018.
\newblock Variational Autoencoders for Collaborative Filtering.
\newblock In \emph{Proceedings of the 2018 World Wide Web Conference}, WWW '18,
  689–698. Republic and Canton of Geneva, CHE: International World Wide Web
  Conferences Steering Committee.
\newblock ISBN 9781450356398.

\bibitem[{London and Sandler(2019)}]{pacbayes}
London, B.; and Sandler, T. 2019.
\newblock {B}ayesian Counterfactual Risk Minimization.
\newblock In Chaudhuri, K.; and Salakhutdinov, R., eds., \emph{Proceedings of
  the 36th International Conference on Machine Learning}, volume~97 of
  \emph{Proceedings of Machine Learning Research}, 4125--4133. PMLR.

\bibitem[{Ma, Wang, and Narayanaswamy(2019)}]{iml}
Ma, Y.; Wang, Y.-X.; and Narayanaswamy, B. 2019.
\newblock Imitation-Regularized Offline Learning.
\newblock In Chaudhuri, K.; and Sugiyama, M., eds., \emph{Proceedings of the
  Twenty-Second International Conference on Artificial Intelligence and
  Statistics}, volume~89 of \emph{Proceedings of Machine Learning Research},
  2956--2965. PMLR.

\bibitem[{Malkov and Yashunin(2020)}]{hnsw}
Malkov, Y.~A.; and Yashunin, D.~A. 2020.
\newblock Efficient and Robust Approximate Nearest Neighbor Search Using
  Hierarchical Navigable Small World Graphs.
\newblock \emph{IEEE Trans. Pattern Anal. Mach. Intell.}, 42(4): 824–836.

\bibitem[{Masrani, Le, and Wood(2019)}]{tvo}
Masrani, V.; Le, T.~A.; and Wood, F. 2019.
\newblock The Thermodynamic Variational Objective.
\newblock In Wallach, H.; Larochelle, H.; Beygelzimer, A.; d\textquotesingle
  Alch\'{e}-Buc, F.; Fox, E.; and Garnett, R., eds., \emph{Advances in Neural
  Information Processing Systems}, volume~32. Curran Associates, Inc.

\bibitem[{Mikolov et~al.(2013)Mikolov, Chen, Corrado, and
  Dean}]{mikolov2013efficient}
Mikolov, T.; Chen, K.; Corrado, G.; and Dean, J. 2013.
\newblock Efficient estimation of word representations in vector space.
\newblock \emph{arXiv preprint arXiv:1301.3781}.

\bibitem[{Paszke et~al.(2019)Paszke, Gross, Massa, Lerer, Bradbury, Chanan,
  Killeen, Lin, Gimelshein, Antiga, Desmaison, Kopf, Yang, DeVito, Raison,
  Tejani, Chilamkurthy, Steiner, Fang, Bai, and Chintala}]{torch}
Paszke, A.; Gross, S.; Massa, F.; Lerer, A.; Bradbury, J.; Chanan, G.; Killeen,
  T.; Lin, Z.; Gimelshein, N.; Antiga, L.; Desmaison, A.; Kopf, A.; Yang, E.;
  DeVito, Z.; Raison, M.; Tejani, A.; Chilamkurthy, S.; Steiner, B.; Fang, L.;
  Bai, J.; and Chintala, S. 2019.
\newblock PyTorch: An Imperative Style, High-Performance Deep Learning Library.
\newblock In Wallach, H.; Larochelle, H.; Beygelzimer, A.; d\textquotesingle
  Alch\'{e}-Buc, F.; Fox, E.; and Garnett, R., eds., \emph{Advances in Neural
  Information Processing Systems 32}, 8024--8035. Curran Associates, Inc.

\bibitem[{Rappaz, McAuley, and Aberer(2021)}]{twitch}
Rappaz, J.; McAuley, J.; and Aberer, K. 2021.
\newblock \emph{Recommendation on Live-Streaming Platforms: Dynamic
  Availability and Repeat Consumption}, 390–399.
\newblock New York, NY, USA: Association for Computing Machinery.
\newblock ISBN 9781450384582.

\bibitem[{Rawat et~al.(2019)Rawat, Chen, Yu, Suresh, and
  Kumar}]{rawat2019sampled}
Rawat, A.~S.; Chen, J.; Yu, F. X.~X.; Suresh, A.~T.; and Kumar, S. 2019.
\newblock Sampled softmax with random fourier features.
\newblock \emph{Advances in Neural Information Processing Systems}, 32.

\bibitem[{Rendle et~al.(2009)Rendle, Freudenthaler, Gantner, and
  Schmidt-Thieme}]{bpr}
Rendle, S.; Freudenthaler, C.; Gantner, Z.; and Schmidt-Thieme, L. 2009.
\newblock BPR: Bayesian Personalized Ranking from Implicit Feedback.
\newblock In \emph{Proceedings of the Twenty-Fifth Conference on Uncertainty in
  Artificial Intelligence}, UAI '09, 452–461. Arlington, Virginia, USA: AUAI
  Press.
\newblock ISBN 9780974903958.

\bibitem[{Robbins and Monro(1951)}]{rm_sgd}
Robbins, H.; and Monro, S. 1951.
\newblock {A Stochastic Approximation Method}.
\newblock \emph{The Annals of Mathematical Statistics}, 22(3): 400 -- 407.

\bibitem[{Ruder(2016)}]{sgd}
Ruder, S. 2016.
\newblock An overview of gradient descent optimization algorithms.
\newblock \emph{arXiv preprint arXiv:1609.04747}.

\bibitem[{Sakhi et~al.(2020)Sakhi, Bonner, Rohde, and Vasile}]{blob}
Sakhi, O.; Bonner, S.; Rohde, D.; and Vasile, F. 2020.
\newblock BLOB: A Probabilistic Model for Recommendation That Combines Organic
  and Bandit Signals.
\newblock In \emph{Proceedings of the 26th ACM SIGKDD International Conference
  on Knowledge Discovery \&; Data Mining}, KDD '20, 783–793. New York, NY,
  USA: Association for Computing Machinery.
\newblock ISBN 9781450379984.

\bibitem[{Sakhi, Faury, and Vasile(2020)}]{dro2}
Sakhi, O.; Faury, L.; and Vasile, F. 2020.
\newblock Improving Offline Contextual Bandits with Distributional Robustness.
\newblock arXiv:2011.06835.

\bibitem[{Steck(2020)}]{esae}
Steck, H. 2020.
\newblock Autoencoders that don't overfit towards the Identity.
\newblock In Larochelle, H.; Ranzato, M.; Hadsell, R.; Balcan, M.; and Lin, H.,
  eds., \emph{Advances in Neural Information Processing Systems 33: Annual
  Conference on Neural Information Processing Systems 2020, NeurIPS 2020,
  December 6-12, 2020, virtual}.

\bibitem[{Su et~al.(2020)Su, Dimakopoulou, Krishnamurthy, and Dud{\'\i}k}]{drc}
Su, Y.; Dimakopoulou, M.; Krishnamurthy, A.; and Dud{\'\i}k, M. 2020.
\newblock Doubly robust off-policy evaluation with shrinkage.
\newblock In \emph{International Conference on Machine Learning}, 9167--9176.
  PMLR.

\bibitem[{Sutton and Barto(2018)}]{rl}
Sutton, R.~S.; and Barto, A.~G. 2018.
\newblock \emph{Reinforcement learning: An introduction}.
\newblock MIT press.

\bibitem[{Swaminathan and Joachims(2015{\natexlab{a}})}]{poem}
Swaminathan, A.; and Joachims, T. 2015{\natexlab{a}}.
\newblock Counterfactual Risk Minimization: Learning from Logged Bandit
  Feedback.
\newblock In Bach, F.; and Blei, D., eds., \emph{Proceedings of the 32nd
  International Conference on Machine Learning}, volume~37 of \emph{Proceedings
  of Machine Learning Research}, 814--823. Lille, France: PMLR.

\bibitem[{Swaminathan and Joachims(2015{\natexlab{b}})}]{snips}
Swaminathan, A.; and Joachims, T. 2015{\natexlab{b}}.
\newblock The Self-Normalized Estimator for Counterfactual Learning.
\newblock In \emph{NIPS}.

\bibitem[{Tanielian and Vasile(2019)}]{negative_sampling}
Tanielian, U.; and Vasile, F. 2019.
\newblock Relaxed Softmax for PU Learning.
\newblock In \emph{Proceedings of the 13th ACM Conference on Recommender
  Systems}, RecSys '19, 119–127. New York, NY, USA: Association for Computing
  Machinery.
\newblock ISBN 9781450362436.

\bibitem[{Wan and McAuley(2018)}]{gr1}
Wan, M.; and McAuley, J.~J. 2018.
\newblock Item recommendation on monotonic behavior chains.
\newblock In Pera, S.; Ekstrand, M.~D.; Amatriain, X.; and O'Donovan, J., eds.,
  \emph{Proceedings of the 12th {ACM} Conference on Recommender Systems, RecSys
  2018, Vancouver, BC, Canada, October 2-7, 2018}, 86--94. {ACM}.

\bibitem[{Wan et~al.(2019)Wan, Misra, Nakashole, and McAuley}]{gr2}
Wan, M.; Misra, R.; Nakashole, N.; and McAuley, J.~J. 2019.
\newblock Fine-Grained Spoiler Detection from Large-Scale Review Corpora.
\newblock In Korhonen, A.; Traum, D.~R.; and M{\`{a}}rquez, L., eds.,
  \emph{Proceedings of the 57th Conference of the Association for Computational
  Linguistics, {ACL} 2019, Florence, Italy, July 28- August 2, 2019, Volume 1:
  Long Papers}, 2605--2610. Association for Computational Linguistics.

\bibitem[{Wang, Agarwal, and Dud{\i}k(2017)}]{switch}
Wang, Y.-X.; Agarwal, A.; and Dud{\i}k, M. 2017.
\newblock Optimal and adaptive off-policy evaluation in contextual bandits.
\newblock In \emph{International Conference on Machine Learning}, 3589--3597.
  PMLR.

\bibitem[{Williams(1992)}]{REINFORCE}
Williams, R.~J. 1992.
\newblock Simple statistical gradient-following algorithms for connectionist
  reinforcement learning.
\newblock \emph{Machine Learning}, 8(3): 229--256.

\end{thebibliography}

\end{document}